\documentclass[lettersize,journal]{IEEEtran}
\usepackage{amsmath,amssymb,amsfonts}
\usepackage{cleveref}
\usepackage{algorithmic}
\usepackage{algorithm}
\usepackage{array}
\usepackage[caption=false,font=normalsize,labelfont=sf,textfont=sf]{subfig}
\usepackage{textcomp}
\usepackage{stfloats}
\usepackage{url}
\usepackage{verbatim}
\usepackage{graphicx}
\usepackage{xcolor} 
\usepackage{bm}
\usepackage{multirow} 
\usepackage{diagbox} 
\usepackage{booktabs}
\usepackage{threeparttable}
\usepackage{authblk}
\usepackage{braket}
\usepackage{xspace}
\usepackage{mhchem}
\def\BibTeX{{\rm B\kern-.05em{\sc i\kern-.025em b}\kern-.08em
    T\kern-.1667em\lower.7ex\hbox{E}\kern-.125emX}}
\usepackage{balance}

\iffalse

\else

\fi

\Crefname{figure}{Fig.}{Figs.}
\Crefname{equation}{Eq.}{Eqs.}
\Crefname{algorithm}{Alg.}{Algs.}
\Crefname{table}{Tab.}{Tabs.}

\begin{document}
\title{QSRA: A QPU Scheduling and Resource Allocation Approach for Cloud-Based Quantum Computing}
\author{Binhan Lu, Zhaoyun Chen, Yuchun Wu
\thanks{This work was supported by the National Key Research and Development Program of China (Grant No. 2023YFB4502500).
(Corresponding author: Zhaoyun Chen)
Binhan Lu and Yuchun Wu are with the School of Physics, University of Science and Technology of China, Hefei, Anhui, China (E-mail: sheldonl@mail.ustc.edu.cn; wuyuchun@ustc.edu.cn).
Zhaoyun Chen is with the Institute of Artificial Intelligence, Hefei Comprehensive National Science Center, Hefei, Anhui, China (E-mail: chenzhaoyun@iai.ustc.edu.cn).
}
}

\markboth{Journal of \LaTeX\ Class Files,~Vol.~18, No.~9, September~2020}%
{How to Use the IEEEtran \LaTeX \ Templates}

\maketitle

\begin{abstract}
Quantum cloud platforms, which rely on Noisy Intermediate-Scale Quantum (NISQ) devices, face significant challenges in efficiently managing quantum programs. This paper proposes a QPU Scheduling and Resource Allocation (QSRA) approach to address these challenges. QSRA enhances qubit utilization and reduces turnaround time by adapting CPU scheduling techniques to Quantum Processing Units (QPUs). It incorporates a subroutine for qubit allocation that takes into account qubit quality and connectivity, while also merging multiple quantum programs to further optimize qubit usage. Our evaluation of QSRA against existing methods demonstrates its effectiveness in improving both qubit utilization and turnaround time.
\end{abstract}


\section{Introduction}
Quantum computing can outperform classical methods in many areas. 
The maintenance of quantum computers is a high-cost endeavor. 
Vendors aim to provide an easy-to-use interface for end-users to conveniently access quantum computers through cloud services, 
simplifying backend management for users. 

However, challenges remain due to Noisy Intermediate Scale Quantum (NISQ) chips \cite{preskill2018quantum}, which have limited qubit coherence and error-prone gates, 
and lack fault tolerance \cite{devitt2013quantum}. 
Although mapping methods can optimize qubit fidelity~\cite{deng2020codar}, they are unsuitable for cloud platforms as they fail to schedule multiple programs simultaneously, limiting resource utilization.
Firstly, as the scale of quantum chips increases, their computing resources (the number of qubits) exceed the number required for a single user-submitted task.
Without a scheduling algorithm, the chip can only process the first task in the queue at a time, leading to significant task backlogs, 
low throughput for the cloud platform, and many idle qubits on the chip.
Secondly, cloud platforms must handle high-frequency quantum program submissions from users, making a scheduling algorithm for computing resources essential. 
Existing methods like QHSP~\cite{niu2021enabling} have made strides in multi-program scheduling, with similar approaches also suggested by~\cite{das2019case,niu2022parallel,2021QuCloud}.
However, these lack task prioritization or rely solely on CZ gate density, 
neglecting submission times and leading to longer waiting periods. Wu et al.~\cite{wu2024reducing} proposed RedLent, 
which prioritizes scheduling based on a linear combination of waiting time, qubit count, and execution time. 
However, choosing suboptimal parameters can significantly prolong program turnaround times. 
Meanwhile, Ohkura et al. \cite{ohkura2021simultaneous} introduced XtalkAw, which uses spatial buffers between parallel programs to reduce crosstalk, 
but this approach lowers chip resource utilization.

In this paper, we propose a QPU Scheduling and Resource Allocation (QSRA) Approach for quantum cloud platforms, consisting of three modules: 
task scheduling, resource allocation, and quantum program merging. 
Task scheduling leverages proven CPU scheduling techniques \cite{harki2020cpu}. 
To minimize crosstalk while maintaining high qubit utilization, circuits are merged or separately allocated based on execution time. 
During qubit allocation, a connection ratio is calculated to avoid elongated regions that could increase execution time and reduce utilization.
This approach maximizes quantum chip resource utilization and boosts cloud platform throughput.

\section{Background}
\subsection{Error Mechanism}
Quantum systems are highly sensitive to decoherence \cite{ithier2005decoherence}. 
Decoherence error increases exponentially over time.
Meanwhile, crosstalk, resulting from unintended qubit coupling leads to increased errors during the parallel execution of near quantum gates \cite{krinner2020benchmarking}.
Finally, measurement errors play an important role \cite{nation2021scalable}.
\subsection{QPU Scheduling Techniques}
Various studies have examined CPU scheduling techniques based on process management and performance evaluation \cite{harki2020cpu}.
First Come First Serve (FCFS) handles processes in the order they arrive. 
FCFS is simple and fair but can cause long wait times for short jobs. 
Shortest Job First (SJF) prioritizes tasks with the least processing requirements.
SJF reduces waiting times for short tasks but may starve longer ones.
Shortest Remaining Time First (SRTF) is a preemptive algorithm that selects the process with the least remaining execution time.
SRTF minimizes average waiting times but risks starvation for longer tasks.
Round-robin (RR) allocates a fixed time slice for each process, cycling through them if not completed.
RR offers fairness and reasonable response times but may incur high context-switching overhead and increased waiting times for short tasks. 
A multilevel feedback Queue (MFQ) dynamically adjusts priorities and allows task movement between queues, using aging techniques to prevent starvation. 
It effectively balances the execution of short and long tasks, but its complexity in implementation, potential overhead, and the need for careful tuning are notable drawbacks.
Highest Response Ratio First (HRRF) is a non-preemptive algorithm that prioritizes based on the response ratio 
$t_\text{res}=(t_\text{wait}+t_\text{ser})/t_\text{ser}$.
HRRF balances execution between short and long processes while preventing starvation but potentially reducing efficiency for shorter tasks.
Selecting the appropriate scheduling algorithm is essential for optimizing performance in different situations.

\begin{figure*}[htbp]
\begin{minipage}[t]{0.35\linewidth}
\centering
\includegraphics[width=\textwidth]{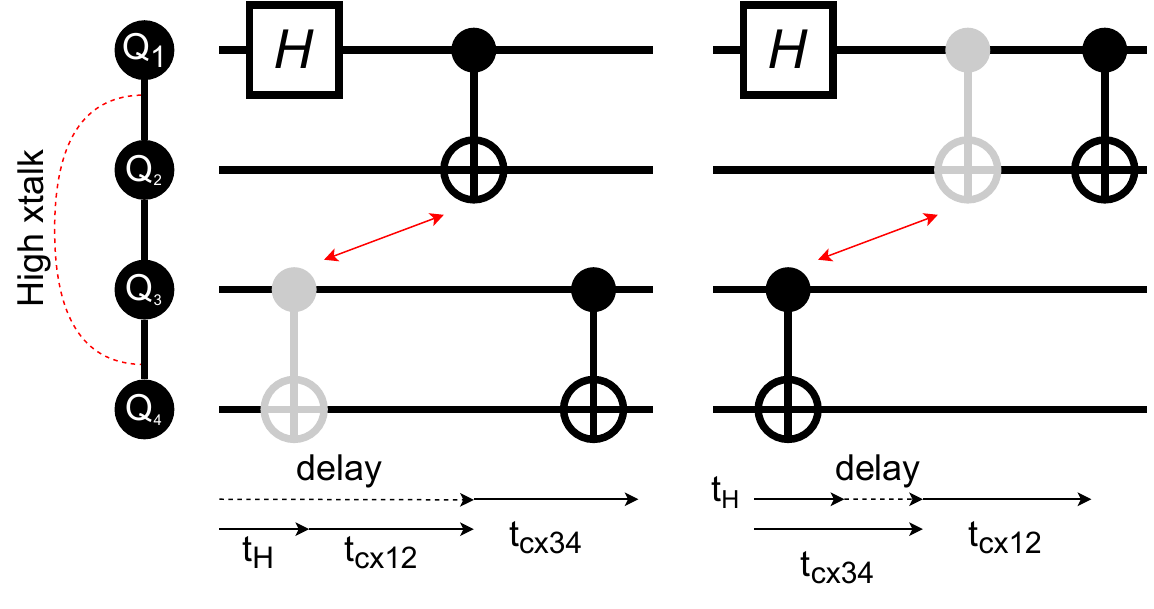}\\
(a)
\end{minipage}
\begin{minipage}[t]{0.17\linewidth}
\centering
\includegraphics[width=\textwidth]{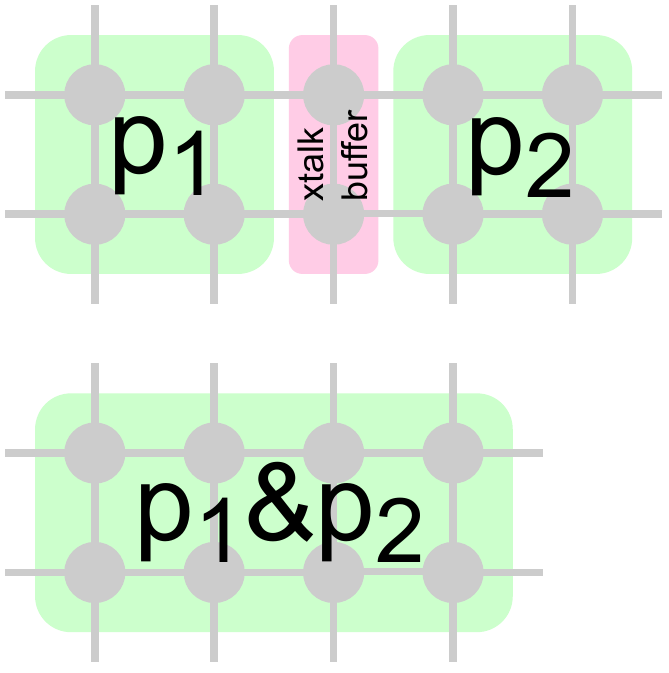}\\
(b)
\end{minipage}
\begin{minipage}[t]{0.45\linewidth}
\centering
\includegraphics[width=\textwidth]{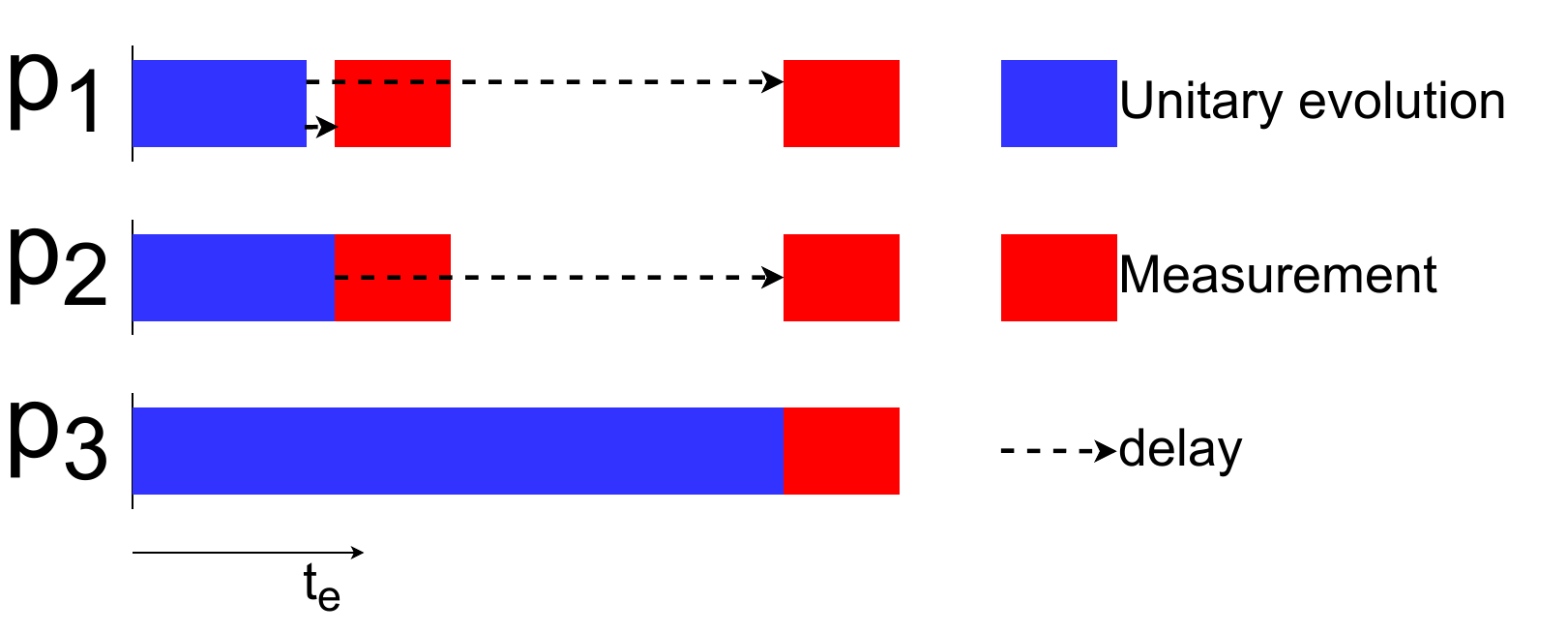}\\
(c)
\end{minipage}
\caption{
(a) Within the same program, if high crosstalk occurs, the affected two-qubit gates must be temporally separated. The delay for $CX_{12}$ is shorter than that for $CZ_{34}$, i.e., $t_{cx} - t_H < t_{cx} + t_H$. 
To minimize decoherence noise, selecting the shorter delay is preferable, framing this as an optimization problem.
(b) If $p_1$ and $p_2$ are not merged, their execution is independent, and gate timing information between them is not shared, 
making it impossible to perform crosstalk suppression as described in (a). 
As a result, a buffer must be left on the chip to prevent qubits of $p_{1,2}$ from being neighbors of qubits in $p_{2,1}$. 
However, if the programs are merged into $p_1\&p_2$, gate timing information is shared, allowing for crosstalk suppression optimization, 
and no buffer is needed between the qubits of the two programs.
(c) The $t_e$ of $p_1$ and $p_2$ are similar but differ greatly from $p_3$, merging 
$p_1$ or $p_2$ with $p_3$ would extend their execution time, increasing their decoherence error.
}
\label{fig merge}
\end{figure*}

\section{Methodology}
\subsection{QPU Scheduling}
Suppose the current job queue $\bm{q}$ is comprised of $K$ quantum programs to be executed, i.e. $\bm{q}=\{p_1,\cdots,p_k\}$. 
Each job $p_i$ has a tuple $(n_i,s_i,t_i)$, where $n_i,s_i$ and $t_i$ denote the qubit number, measurement shot number, and submission time, respectively.
For a quantum computer, $t_e$ includes execution time along with measurement and initialization.
For a quantum program, $t_E = s_i t_e$ represents the execution and result collection time. 
$t_E$ is used for SJF scheduling, while $t_i$ is used for FCFS.
For RR and SRTF, time slices cannot be set during unitary evolution but must occur between measurements and initialization.

With different scheduling methods, the priority of quantum programs in the queue will be reordered, and those at the front will be executed first.
We will evaluate each scheduling scheme using weighted turnaround time and throughput. 
Turnaround time is defined as the total time from submission to completion, $t_\text{turn} = t_\text{comp} - t_\text{sub}$.
Weighted turnaround time, also known as normalized turnaround time, is the ratio of the turnaround time to the process's service time (execution time) $t_\text{wt}=t_\text{turn}/t_E$. 
It helps compare the efficiency of scheduling algorithms with different service times.

Classical CPU scheduling only needs to consider program service time, but quantum chips have an additional attribute: qubit count. 
If the number of qubits required by a program $n_i$ is less than the total qubits $N$ available on the chip, 
the chip allows another program, with a qubit requirement less than $N - n_i$, to execute on the unoccupied qubits.
Based on qubit count, we introduce quantum versions of SJF and HRRF scheduling schemes: QSJF and QHRRF. 
We define $\eta = n_i / N$ as the qubit utilization rate for a single program. 
In QSJF, program size is evaluated by $\eta t_\text{ser}$, prioritizing the smallest program size instead of the shortest service time. 
In QHRRF, the response time is calculated as $t_\text{qres} = (t_\text{wait} + \eta t_\text{ser}) / t_\text{ser}$.

\subsection{Qubit Resource Allocation \& Circuit Merging}
After sorting, we sequentially select the tasks at the front of the queue $\bm{q}$ and allocate qubit resources for them.
As mentioned earlier, quantum chips allow multiple programs to run simultaneously. 
In this section, we will discuss how to allocate qubit resources efficiently for multiple programs on the chip.

\begin{figure}[htbp]
\hspace{0.25cm} 
\begin{minipage}[t]{0.35\linewidth}
\centering
\includegraphics[width=\textwidth]{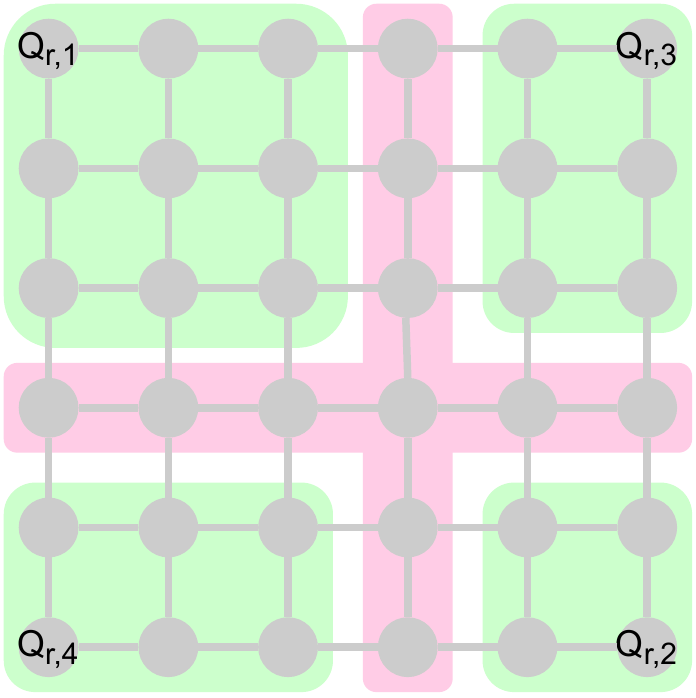}\\
(a)
\end{minipage}
\hspace{1.2cm} 
\begin{minipage}[t]{0.45\linewidth}
\centering
\includegraphics[width=\textwidth]{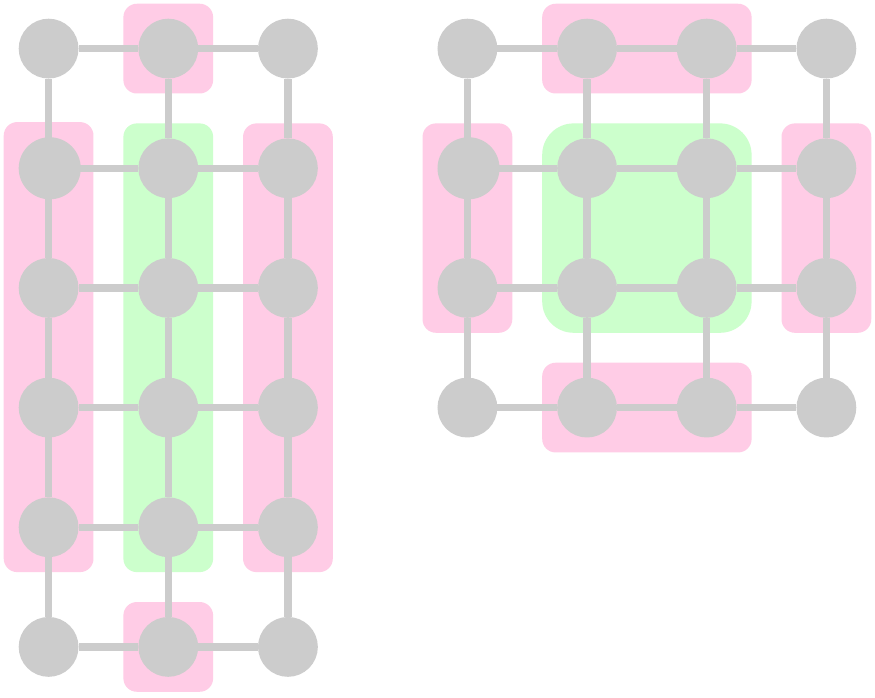}\\
(b)
\end{minipage}
\caption{
(a) The root nodes of the four programs to be allocated are placed in the four corners because the four corners maximize the sum of distances between the root qubits on this chip.
The red region represents the buffer qubits.
(b) On the left, a four-qubit program is allocated in a linear region with 3 internal edges and 10 external edges, giving a ratio of $r_i/r_a=3/(10+3)=0.23$. 
On the right, there are 4 internal edges and 8 external edges, resulting in a ratio of $4/(4+8)=0.33$. 
Consequently, the buffer region on the left is larger than on the right, indicating that interconnectivity in the left region is sparser than in the right.
}
\label{fig alloc}
\end{figure}

\subsubsection{Circuit Merging}
Resulting of crosstalk, gate fidelity will be degraded during the parallel execution of multiple quantum gates \cite{krinner2020benchmarking}.
In \cite{murali2020software,hua2022cqc}, several crosstalk suppression strategies have been proposed. 
These strategies optimize gate sequencing within a quantum program, as shown in \Cref{fig merge}(a), but require gate timing information, which is unavailable between different quantum programs.
If two programs are spatially close and their timing is uncertain, qubits belonging to $p_{1,2}$ may experience crosstalk with qubits belonging to $p_{2,1}$ as in \Cref{fig merge}(b).
One viable approach is to merge two programs for gate timing scheduling, 
which clarifies the timing relationships between the quantum gates of the two programs. 
This allows for the application of existing strategies to suppress crosstalk.
However, if there is a significant disparity in execution times between the two quantum programs, 
the shorter program may be forced to extend its execution time as in \Cref{fig merge}(c), 
leading to increased decoherence errors. Additionally, the qubits that could have been released early for the next program's execution remain occupied.

Therefore, if the total number of qubits required by the first $p$ programs in the queue, $\sum_{p}n_p<N$, 
then the programs should be grouped based on $t_e$, and each group should be allocated and compiled as a single program. 
Programs in the same group with similar $t_e$.
The priority of the merged program is equal to the highest priority among the programs before the merge.
Additionally, qubits assigned to different groups should not be neighboring qubits on the chip.


\begin{table*}[htbp]
\centering
\begin{threeparttable}
\centering
\caption{
Comparison Results
}
\begin{tiny}
\begin{tabular}{cccccccccc|ccccc}
\toprule
\multicolumn{2}{c}{\diagbox{Metrics}{Scheduling Techniques}}&FCFS&SJF&QSJF&SRTF&RR&MFQ&HRRF&QHRRF&QSRA&CODAR&QHSP&RedLent&XtalkAw\\
\midrule
throughput$(1/s)$&$\lambda=5$&$5.11$&$5.24$&$\textbf{6.22}$&$\textbf{6.05}$&$4.31$&$4.20$&$5.15$&$\textbf{5.90}$&$\textbf{6.12}$&$\textbf{5.01}$&$4.98$&$\textbf{6.63}$&$1.54$\\
                &$\lambda=20$&$17.3$&$22.1$&$\textbf{24.5}$&$\textbf{28.4}$&$16.7$&$20.8$&$21.2$&$\textbf{23.5}$&$\textbf{19.5}$&$6.06$&$\textbf{10.90}$&$8.77$&$\textbf{12.1}$\\
                &$\lambda=50$&$42.5$&$54.1$&$\textbf{65.1}$&$\textbf{68.8}$&$34.6$&$36.9$&$53.3$&$\textbf{55.4}$&$\textbf{46.5}$&$5.83$&$17.9$&$\textbf{40.7}$&$\textbf{33.3}$\\
                &$\lambda=100$&$80.7$&$92.7$&$\textbf{115}$&$\textbf{128}$&$74.5$&$86.5$&$103$&$\textbf{105}$&$\textbf{87.6}$&$6.21$&$18.8$&$\textbf{79.7}$&$\textbf{64.4}$\\
                &$\lambda=500$&$85.1$&$82.3$&$\textbf{133}$&$\textbf{123}$&$84.4$&$76.1$&$113$&$\textbf{125}$&$\textbf{127}$&$4.20$&$15.8$&$\textbf{104}$&$\textbf{88.4}$\\
utilization$(\%)$&$\lambda=5$&$30.1$&$29.6$&$30.4$&$33.1$&$\textbf{35.3}$&$33.7$&$\textbf{34.2}$&$\textbf{38.5}$&$\textbf{61.1}$&$10.6$&$49.8$&$\textbf{66.3}$&$\textbf{50.2}$\\
                &$\lambda=20$&$60.4$&$76.8$&$\textbf{90.5}$&$\textbf{77.7}$&$61.2$&$56.4$&$71.7$&$\textbf{83.3}$&$\textbf{70.5}$&$6.20$&$49.7$&$\textbf{77.4}$&$\textbf{61.4}$\\
                &$\lambda=50$&$70.3$&$\textbf{80.5}$&$\textbf{92.2}$&$70.6$&$54.4$&$66.8$&$53.4$&$\textbf{85.3}$&$\textbf{83.5}$&$8.00$&$\textbf{59.2}$&$\textbf{88.3}$&$34.1$\\
                &$\lambda=100$&$67.3$&$72.6$&$\textbf{86.3}$&$\textbf{75.6}$&$41.5$&$63.3$&$64.0$&$\textbf{87.1}$&$\textbf{82.0}$&$2.00$&$\textbf{68.6}$&$\textbf{85.5}$&$41.3$\\
                &$\lambda=500$&$60.0$&$70.2$&$\textbf{90.5}$&$\textbf{78.6}$&$52.3$&$73.1$&$53.9$&$\textbf{92.1}$&$\textbf{85.4}$&$4.23$&$\textbf{58.6}$&$\textbf{83.5}$&$41.8$\\
weighted turnaround time&$\lambda=5$&$\textbf{1.29}$&$\textbf{1.01}$&$1.98$&$1.63$&$1.53$&$1.43$&$1.51$&$\textbf{1.09}$&$\textbf{1.61}$&$\textbf{4.10}$&$4.98$&$\textbf{1.66}$&$5.94$\\
                        &$\lambda=20$&$\textbf{1.63}$&$4.06$&$4.95$&$4.77$&$2.61$&$\textbf{2.56}$&$2.71$&$\textbf{1.83}$&$\textbf{1.73}$&$26.0$&$49.3$&$\textbf{2.67}$&$\textbf{10.6}$\\
                        &$\lambda=50$&$\textbf{2.78}$&$5.87$&$7.92$&$5.73$&$\textbf{3.34}$&$3.66$&$3.53$&$\textbf{2.85}$&$\textbf{2.86}$&$70.8$&$59.0$&$\textbf{6.60}$&$\textbf{34.0}$\\
                        &$\lambda=100$&$4.77$&$6.24$&$8.86$&$9.05$&$\textbf{4.41}$&$\textbf{4.40}$&$4.63$&$\textbf{3.82}$&$\textbf{3.87}$&$120$&$86.1$&$\textbf{12.6}$&$\textbf{41.4}$\\
                        &$\lambda=500$&$8.97$&$9.72$&$15.6$&$14.7$&$\textbf{8.41}$&$\textbf{7.63}$&$8.63$&$\textbf{7.82}$&$\textbf{7.92}$&$801$&$586$&$\textbf{20.7}$&$\textbf{441}$\\
PST$(\%)$               &             &     &     &       &       &      &      &      &      &$\textbf{75.7}$&$\textbf{78.4}$&$68.2$&$33.0$&$\textbf{72.4}$\\
\bottomrule
\end{tabular}\label{tab res}
\begin{tablenotes}
\item[i] The left 8 columns show the performances of eight different QPU scheduling techniques across throughput, utilization, and weighted turnaround time metrics. The right 5 columns display the performances of QSRA and the CODAR, QHSP, RedLent, and XtalkAw baselines on these same metrics, with the final row showing performance based on the PST metric. 
\item[ii] The table compares the performances of throughput, utilization, and weighted turnaround time at Poisson rates of 5, 20, 50, 100, and 500. 
The top three performing methods in each row are highlighted in \textbf{bold}.
\item[iii] Higher throughput and utilization are better, while lower weighted turnaround time and higher PST indicate improved performance.
\end{tablenotes}
\end{tiny}
\end{threeparttable}
\end{table*}

\subsubsection{Qubit Allocation}
In this section, we discuss how to allocate qubit resources for multiple quantum programs. 
First, for the coupling graph of the chip, a root qubit $Q_r$ is determined, and qubits are gradually added around this root until the number of qubits matches the program's requirement. 
Second, as in \Cref{fig alloc}(a), if more than one program requires allocation, to avoid crosstalk, the root qubit $Q_{r,i}$ of each program is placed in a location that maximizes the distance from the previous root qubit $Q_{r,i-1}$. 
Third, as shown in \Cref{fig alloc}(b), when allocating qubits to each program, we calculate the ratio $r_i/r_a$, 
where $r_i$ is the number of edges within the allocation region and $r_a$ is the total number of edges connected to that region. 
A low ratio indicates sparse connectivity, which increases swap gate \cite{li2019tackling} during compilation. 
Therefore, when adding qubits, we should select the qubit that maximizes the ratio.
Fourth, When adding qubits, if multiple qubits correspond to the same $r_i/r_a$, 
We calculate $E_Q=(1-\exp(t_e/T_Q))E_\text{meas}$, 
where $T_Q$ is the coherence time of $Q$, 
$(1-\exp(t_e/T_Q))$ represents the decoherence errors during the duration $t_e$, 
and $E_\text{meas}$ is the measurement error of qubit $Q$. 
We select the qubit with the minimum $E_Q$ for the next allocation.
Finally, if the remaining qubits available for program $p_i$ are all neighboring qubits of program $p_j$ (causing crosstalk), 
the program with the lower priority between $p_i$ and $p_j$ is returned to the queue for the next execution cycle. 
If $p_i$ and $p_j$ are merged programs, the subprogram with the lowest priority in $p_i$ or $p_j$ will be returned to the queue. 
This process is repeated until there is no crosstalk.

\section{Results}
\subsection{Metrics}
\textbf{Probability of a Successful Trial (PST)}, 
the ratio of the number of successful trials to the total number of trials performed \cite{das2019case}.
\textbf{Weighted turnaround time}, 
shorter turnaround times lead to faster results for users, improving cloud platform throughput. 
\textbf{Utilization},
maximizing qubit utilization can increase cloud platform throughput.

The user submissions on the cloud platform are modeled by a Poisson process \cite{daley2003basic}, with the arrival rate reflecting submission frequency.
\subsection{Chip Model, Baselines \& Benchmarks}
We used IBM's IBM Fez chip and qubit data as the physical model for our simulations.
First, we identify the best-performing QPU scheduling technique for integration into QSRA. Then, we compare QSRA with these baselines:
(i) CODAR \cite{deng2020codar} for single-program scheduling, 
(ii) QHSP \cite{niu2021enabling} for multi-program scheduling, 
(iii) Wu et al.'s RedLent manual priority-setting \cite{wu2024reducing}, 
and (iv) Ohkura's XtalkAw multi-program scheduling that accounts for crosstalk \cite{ohkura2021simultaneous}.
We use the quantum circuits from \cite{soeken2012revkit} as benchmarks to compare different QPU scheduling methods. 
Additionally, to evaluate the performance against current baselines, 
we use the variational quantum algorithm (VQE) circuits for computing the ground state energies of \ce{H2}, \ce{N2}, \ce{LiH}, \ce{H2O}, \ce{NH3}, and \ce{CO2}.

\subsection{QPU Scheduling Methods Comparison}
From \Cref{tab res}, we observe that QSJF, SRTF, and QHRRF consistently rank in the top three for throughput and utilization, especially at higher Poisson rates. 
This is because SJF, SRTF, QSJF, QHRRF, and HRRF prioritize shorter jobs, allowing the QPU to complete more tasks and thereby increase throughput. 
Additionally, QSJF and QHRRF give precedence to jobs with fewer qubits, enabling greater parallelization and improving utilization. 
However, SJF, SRTF and QSJF may delay longer jobs, leading to prolonged wait times for programs with higher $T_E$, which results in a longer average weighted turnaround time. 
In contrast, HRRF and QHRRF use the highest response ratio metric to balance waiting time and job size effectively. Overall, QHRRF emerges as the optimal QPU scheduling solution.

\subsection{Comparison with Baselines}
From \Cref{tab res}, the QSRA approach with QHRRF scheduling and RedLent demonstrates superior throughput, chip utilization, and weighted turnaround time compared to QHSP and XtalkAw, with CODAR performing the worst. This performance gap is due to QHSP and XtalkAw overlooking task submission timing, while CODAR’s single-job execution constraint lowers utilization. For PST, CODAR ranks highest, closely followed by QSRA, which also outperforms XtalkAw and QHSP, while RedLent exhibits the lowest PST. CODAR’s single-program mapping supports optimal resource allocation, enhancing its PST performance.

\begin{figure}[htbp]
\centering
\includegraphics[width=0.24\textwidth]{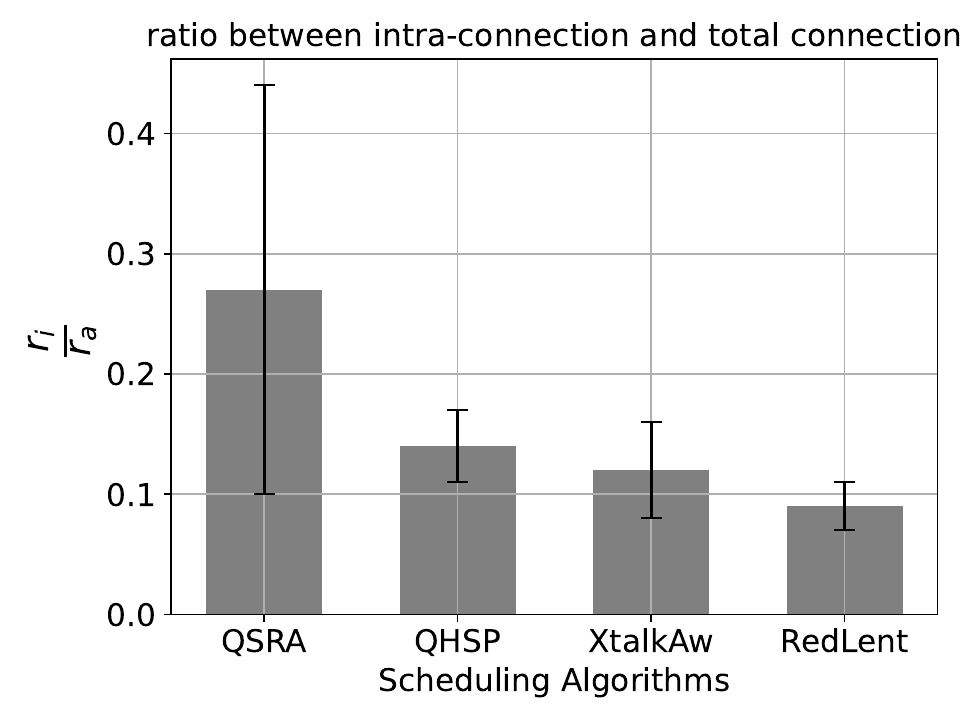}\\
\caption{
The ratio of internal to total connections within the program after qubit resource allocation across different baselines.
}
\label{fig comp}
\end{figure}

To explain the lower performance of the other baselines, we calculated the average $r_i/r_a$, finding RedLent to be the lowest, followed by QHSP and XtalkAw, both below QSRA (see \Cref{fig comp}). While QHSP and XtalkAw account for crosstalk in qubit assignments, they overlook connectivity within the execution area. This oversight increases the need for swap gates, thereby prolonging execution times and heightening decoherence errors. RedLent, which disregards both crosstalk and connectivity, results in the lowest PST among these methods.

By integrating QPU scheduling with resource allocation, QSRA with QHRRF effectively manages real-time task submissions on cloud platforms, achieving high fidelity for quantum programs.

\section{Conclusion}
In this study, we proposed the QSRA approach with the QHRRF scheduling scheme to optimize quantum program execution on cloud platforms. 
By addressing QPU scheduling and resource allocation, QSRA effectively manages task submissions, enhancing throughput, 
chip utilization, and turnaround time. Our results show that QSRA improves resource efficiency while maintaining high fidelity for quantum programs, surpassing existing methods, 
and highlighting the potential of advanced scheduling techniques in cloud-based quantum computing environments.

\bibliographystyle{unsrt}
\bibliography{IEEEexample}


\end{document}